\documentclass[twocolumn,superscriptaddress,floatfix,aps,prd,nofootinbib,showpacs]{revtex4}

\usepackage{graphicx}

\begin{document}

\title{Quasiequilibrium sequences of black-hole--neutron-star binaries
  in general relativity}

\author{Keisuke Taniguchi}
\affiliation{Department of Physics, University of Illinois at
  Urbana-Champaign, Urbana, Illinois 61801, USA}
\author{Thomas W. Baumgarte}
\altaffiliation{Also at: Department of Physics, University of Illinois
  at Urbana-Champaign, Urbana, Illinois 61801, USA}
\affiliation{Department of Physics and Astronomy, Bowdoin College,
  Brunswick, Maine 04011, USA}
\author{Joshua A. Faber}
\affiliation{Department of Physics, University of Illinois at
  Urbana-Champaign, Urbana, Illinois 61801, USA}
\author{Stuart L. Shapiro}
\altaffiliation{Also at: Department of Astronomy and NCSA, University
  of Illinois at Urbana-Champaign, Urbana, Illinois 61801, USA}
\affiliation{Department of Physics, University of Illinois at
  Urbana-Champaign, Urbana, Illinois 61801, USA}

\date{September 15, 2006}

\begin{abstract}
We construct quasiequilibrium sequences of black hole-neutron star
binaries for arbitrary mass ratios by solving the constraint equations
of general relativity in the conformal thin-sandwich decomposition.
We model the neutron star as a stationary polytrope satisfying the
relativistic equations of hydrodynamics, and account for the black
hole by imposing equilibrium boundary conditions on the surface of an
excised sphere (the apparent horizon).  In this paper we focus on
irrotational configurations, meaning that both the neutron star and
the black hole are approximately nonspinning in an inertial frame.
We present results for a binary with polytropic index $n=1$, mass
ratio $M_{\rm irr}^{\rm BH} / M_{\rm B}^{\rm NS} = 5$ and neutron
star compaction $M_{\rm ADM,0}^{\rm NS}/R_0 = 0.0879$, where
$M_{\rm irr}^{\rm BH}$ is the irreducible mass of the black hole,
$M_{\rm B}^{\rm NS}$ the neutron star baryon rest-mass, and
$M_{\rm ADM,0}^{\rm NS}$ and $R_0$ the neutron star
Arnowitt-Deser-Misner mass
and areal radius in isolation, respectively.
Our models represent valid solutions to Einstein's constraint
equations and may therefore be employed as initial data for dynamical
simulations of black hole-neutron star binaries.
\end{abstract}


\maketitle


Coalescing black hole-neutron star (hereafter BHNS) binaries are among
the most promising sources of gravitational waves for laser
interferometers \cite{LIGO,GEO,TAMA,VIRGO}. BHNS mergers may reveal a
wealth of astrophysical information (see e.g.~\cite{v00}), and, along
with mergers of binary neutron stars, are also considered primary
candidates for central engines of short-duration gamma-ray bursts
(SGRBs) \cite{ShibaT06,FaberBST06,PriceR06}.  Recent observations of
several SGRBs localized by the {\it Swift} and HETE-2 satellites in
regions with low star formation strongly suggest that a compact binary
merger scenario for SGRBs is favored over models involving the
collapse of massive stars (see, e.g., \cite{Berge06} and references
cited therein).

Significant effort has gone into the study of binary neutron stars and
binary black holes, which are also promising sources of gravitational
radiation.  Fully relativistic simulations of BHNS binaries have
received far less attention.  Most BHNS calculations to date,
including quasiequilibrium (QE) calculations
\cite{Chand69,Fishb73,LaiRS93,LaiW96,TanigN96,Shiba96,UryuE99,WiggiL00,
IshiiSM05} and dynamical treatments
\cite{Mashh75,CarteL83,Marck83,LeeK99,Lee00,RosswSW04,KobayLPM04},
employ Newtonian gravitation in either some or all aspects of their
formulation.  We have recently launched a new effort to study BHNS
binaries in a fully relativistic framework (see also
\cite{Miller01,SopueSL06}), first by constructing QE models
\cite{BaumgSS04,TanigBFS05} and then by employing them as initial data
in dynamical simulations \cite{FaberBSTR06,FaberBST06}.  So far we
have focused on binaries for which the black hole mass is much
greater than the neutron star mass.  For binaries with such extreme
mass ratios the rotation axis can be taken to pass through the center
of the black hole, and the tidal effects of the neutron star on the
black hole may be ignored.  These approximations simplify the problem
considerably (see \cite{BaumgSS04}).  However, they break down for
binaries containing comparable mass companions.  Such systems are more
suitable as SGRB candidates, because the tidal disruption of the
neutron star by the black hole will occur near or outside the
innermost stable circular orbit.  This disruption may be
necessary to create a gaseous accretion disk around the black hole
capable of generating a SGRB \cite{FaberBST06}.  Gravitational
waves from BHNS binaries of comparable mass are detectable by ground
based laser interferometers like LIGO
(Laser Interferometric Gravitational-wave Observatory), while waves
from systems with extreme mass ratios are much lower in frequency and
require space-borne interferometers like LISA
(Laser Interferometer Space Antenna).


In this paper we describe the construction of QE sequences of BHNS
binaries with companions of comparable mass.  We construct such
binaries by solving the constraint equations of general relativity
together with the relativistic equations of hydrodynamic equilibrium
in a stationary spacetime assuming the presence of an approximate
helical Killing vector (see, e.g., the recent reviews
\cite{Cook00,BaumgS03} as well as Sec.~II of \cite{GourgGTMB01}).
Throughout this paper we adopt geometric units with $G=c=1$, where
$G$ denotes the gravitational constant and $c$ the speed of light.
Latin and Greek indices denote purely spatial and spacetime
components, respectively.

The line element in 3+1 form can be written as
\begin{eqnarray}
  ds^2 &=& g_{\mu \nu} dx^{\mu} dx^{\nu}, \nonumber \\
  &=& -\alpha^2 dt^2 +\gamma_{ij} (dx^i +\beta^i dt)
  (dx^j +\beta^j dt),
\end{eqnarray}
where $\alpha$ is the lapse function, $\beta^i$ the shift vector,
$\gamma_{ij}$ the spatial metric, and $g_{\mu \nu}$ the spacetime
metric.  Einstein's equations can then be split into constraint and
evolution equations for the spatial metric $\gamma_{ij}$.  To
decompose the constraint equations we introduce a conformal rescaling
$\gamma_{ij} = \psi^4 \tilde{\gamma}_{ij}$, where $\psi$ is the
conformal factor and $\tilde{\gamma}_{ij}$ the spatial background
metric.  The Hamiltonian constraint then reduces to
\begin{equation}
  \tilde{\nabla}^2 \psi = -2 \pi \psi^5 \rho + {1 \over 8} \psi
  \tilde{R} + {1 \over 12} \psi^5 K^2 - {1 \over 8} \psi^{-7}
  \tilde{A}_{ij} \tilde{A}^{ij}. \label{eq:hamilton}
\end{equation}
Here $\tilde{\nabla}_i$, $\tilde{R}_{ij}$, and $\tilde{R} = \tilde
\gamma^{ij} \tilde R_{ij}$ denote the covariant derivative, the Ricci
tensor, and the scalar curvature associated with
$\tilde{\gamma}_{ij}$.  We also decompose the extrinsic curvature
$K^{ij}$ into its trace ($K$) and traceless ($\tilde{A}^{ij}$) parts,
$K^{ij}=\psi^{-10} \tilde{A}^{ij} +\gamma^{ij} K/3$.

In the conformal thin-sandwich decomposition we express the traceless
part of the extrinsic curvature in terms of the time derivative of the
background metric, $\tilde u_{ij} = \partial_t \tilde \gamma_{ij}$,
and gradients of the shift.  For the construction of equilibrium data
it is reasonable to assume $\tilde u_{ij} = 0$ in a corotating
coordinate system, which yields
\begin{equation} \label{Aij}
  \tilde{A}^{ij} ={\psi^6 \over 2\alpha} \Bigl( \tilde{\nabla}^i
  \beta^j +\tilde{\nabla}^j \beta^i -{2 \over 3} \tilde{\gamma}^{ij}
  \tilde{\nabla}_k \beta^k \Bigr).
\end{equation}
Inserting this into the momentum constraint we then obtain
\begin{eqnarray}
  &&\tilde{\nabla}^2 \beta^i +{1 \over 3} \tilde{\nabla}^i
  (\tilde{\nabla}_j \beta^j) +\tilde{R}^i_j \beta^j \nonumber \\
  &&=16 \pi \alpha \psi^4 j^i +2 \tilde{A}^{ij} \tilde{\nabla}_j
  (\alpha \psi^{-6}) +{4 \over 3} \alpha \tilde{\gamma}^{ij}
  \tilde{\nabla}_j K. \label{eq:momentum}
\end{eqnarray}
It is also reasonable to assume $\partial_t K = 0$, which, from
the evolution equation for the extrinsic curvature, yields
\begin{eqnarray}
  \tilde{\nabla}^2 \alpha &=& 4 \pi \alpha \psi^4 (\rho +S)
  +\alpha \psi^{-8} \tilde{A}_{ij} \tilde{A}^{ij} - \nonumber \\
  && 2 \tilde{\gamma}^{ij} \tilde{\nabla}_i \alpha \tilde{\nabla}_j
  \ln \psi
  +{1 \over 3} \alpha \psi^4 K^2 +\psi^4 \beta^i \tilde{\nabla}_i K.
  \label{eq:trace}
\end{eqnarray}
Before we can solve the above set of gravitational field equations for
$\psi$, $\beta^i$ and $\alpha$, we still need to specify the spatial
background metric $\tilde{\gamma}_{ij}$ and the trace of the
extrinsic curvature $K$.  We choose this background geometry to
describe the Schwarzschild metric expressed in Kerr-Schild coordinates.
Specifically, we choose $\tilde{\gamma}_{ij}=\eta_{ij}+2M_{\rm BH} l_i
l_j/r_{\rm BH}$ and $K=2 \hat{\alpha}_{\rm BH}^3 M_{\rm BH} (1+3M_{\rm
BH}/r_{\rm BH})/r_{\rm BH}^2$.  Here $\eta_{ij}$ is the flat spatial
metric, $M_{\rm BH}$ is the ``bare'' mass of the black hole, $r_{\rm BH}
= (X_{\rm BH}^2 +Y_{\rm BH}^2 +Z_{\rm BH}^2)^{1/2}$ is the coordinate
distance from the black hole center, $l_i =l^i \equiv X_{\rm
BH}^i/r_{\rm BH}$ is the radial vector pointing away from the black hole
center, and $\hat{\alpha}_{\rm BH} \equiv (1+2M_{\rm BH}/r_{\rm
BH})^{-1/2}$ is the lapse function of the Schwarzschild metric in
Kerr-Schild coordinates.  The matter terms on the right-hand side
of Eqs.~(\ref{eq:hamilton}), (\ref{eq:momentum}), and (\ref{eq:trace})
are the projections $\rho \equiv n_{\mu} n_{\nu} T^{\mu \nu}$, $j^i
\equiv -\gamma^i_{\mu} n_{\nu} T^{\mu \nu}$, $S_{ij} \equiv \gamma_{i
\mu} \gamma_{j \nu} T^{\mu \nu}$, and $S \equiv \gamma^{ij} S_{ij}$ of
the stress-energy tensor $T_{\mu \nu}$, where $n_{\mu}$ is the unit
vector normal to the spatial hypersurface.  Assuming an ideal fluid
we have $T_{\mu \nu}= (\rho_0 +\rho_i +P) u_{\mu} u_{\nu} +P g_{\mu
\nu}$, where $u_{\mu}$ is the fluid 4-velocity, $\rho_0$ the baryon
rest-mass density, $\rho_i$ the internal energy density, and $P$ the
pressure.

The elliptic Eqs.~(\ref{eq:hamilton}), (\ref{eq:momentum}), and
(\ref{eq:trace}) require boundary conditions, both at spatial infinity
and on the surface of an excised sphere within the black hole
interior.  At spatial infinity, where the metric becomes
asymptotically flat in an inertial frame, we impose the exact boundary
conditions, and on the excision surface
(apparent horizon) we impose the black hole equilibrium boundary
conditions suggested in \cite{CookP04}.  To construct approximately
nonspinning black holes we set the shift according to Eqs.~(39) and
(50) in \cite{CookP04} with $\Omega_r = \Omega_0$ in
their notation.  In the language of \cite{CaudiCGP06} this assignment
corresponds to the ``leading-order approximation'', and we plan to
improve this approximation as outlined there.

In addition to the field Eqs. (\ref{eq:hamilton}),
(\ref{eq:momentum}), and (\ref{eq:trace}) we have to solve the
equations of relativistic hydrodynamics.  For stationary configuration
the relativistic Euler equation can be integrated once to yield 
\begin{equation}
h \alpha \gamma/\gamma_0 ={\rm constant},
\end{equation} 
where $h=(\rho_0 +\rho_i +P)/\rho_0$ is the fluid specific enthalpy,
and $\gamma$ and $\gamma_0$ are Lorentz factors between the fluid, the
rotating frame, and the inertial frame (see Sec.~II.C.~of
\cite{TanigBFS05} for the definitions).  For irrotational fluids the
fluid velocity can be expressed in terms of the gradient of a velocity
potential $\Psi$.  The equation of continuity then becomes
\begin{equation}
(\rho_0/h) \nabla^{\mu} \nabla_{\mu} \Psi +(\nabla^{\mu} \Psi)
\nabla_{\mu} (\rho_0/h) = 0,
\end{equation} 
where $\nabla_{\mu}$ is the covariant derivative associated with
$g_{\mu \nu}$.
We solve
these equations for a polytropic equation of state $P=\kappa
\rho_0^{\Gamma}$, where $\Gamma = 1 + 1/n$ denotes the adiabatic
index, $n$ is the polytropic index and $\kappa$ is a constant. Here,
we focus on the case $n=1$ (i.e., $\Gamma=2$).

We determine the orbital angular velocity by requiring that the
derivative in the $X$ direction of the enthalpy field at the
center of the neutron star be zero, which implies that the total force
balances at the center of the neutron star \cite{TanigBFS05}.  We
confirm that the angular velocity obtained by this method agrees with
that obtained by requiring the enthalpy at two points 
on the neutron star's surface be equal to within one part in
$10^{-5}$ \cite{BaumgSS04}. 

We locate the axis of rotation by requiring that the total linear
momentum $P^i={1 \over 8\pi} \oint_{\infty} K^{ij} dS_j$ vanish
\cite{BowenY80}.  To do so, we first align the axis of rotation with
the $Z$ axis and place the $X$ axis to be the perpendicular line to the
$Z$ axis that passes through the black hole center.  Given the
equatorial symmetry in the problem, the $Z$ component of the momentum
vanishes automatically. With the orbital angular velocity held fixed,
we drive the $Y$ component of the linear momentum toward zero by
adjusting the $X$ coordinate of each companion, keeping their separation
in the $X$ direction unchanged.  To determine their $Y$ coordinates, we
require that the $X$ component of the linear momentum be zero, but only
adjust the $Y$ coordinate of the neutron star to achieve this.  Thus, we
fix the black hole's center to remain on the $X$ axis at $Y=0$.

Our numerical code uses the spectral method {\sc Lorene} library
routines developed by the Meudon relativity group \cite{Lorene}.  The
computational grid is divided into 10 (8) domains for a black hole
(neutron star) and its exterior, and each domain is covered by $N_r
\times N_{\theta} \times N_{\phi}=33 \times 25 \times 24$ collocation
points except for the closest two where 9 (8) domains are covered by
$N_r \times N_{\theta} \times N_{\phi}=25 \times 21 \times 20$
points.


\begin{table}[t]
\caption{Physical parameters for a binary sequence with mass ratio
$M_{\rm irr}^{\rm BH}/M_{\rm B}^{\rm NS}=5$ and neutron star
compaction $M_{\rm ADM, 0}^{\rm NS}/R_0=0.0879$ (where $R_0$ is the
areal radius of the isolated neutron star).  The baryon rest mass, the
ADM mass, and the isotropic coordinate radius of the neutron star in
isolation are $\bar{M}_{\rm B}^{\rm NS}=0.1$, $\bar{M}_{\rm ADM,
0}^{\rm NS}=0.0956$, and $\bar{r}_0=0.990$ ($\kappa =1$).  We
list the binding energy $E_{\rm b}$, total angular momentum $J$,
orbital angular velocity $\Omega$, maximum density parameter
$q_{\rm max}$, mass-shedding indicator $\chi_{\rm min}$, and
fractional difference $\delta M$ between the ADM mass and the Komar
mass.}
\begin{center}
\begin{tabular}{cccccccc} \hline\hline
  $d/M_0$&$E_{\rm b}/M_0$&$J/M_0^2$&
  $\Omega M_0$&$q_{\rm max}$&$\chi_{\rm min}$&$\delta M$ \\ \hline
  20.46 & -7.18(-3) & 0.679 & 1.10(-2) & 5.83(-2) & 0.892 & 7.95(-4) \\
  18.41 & -8.11(-3) & 0.648 & 1.28(-2) & 5.84(-2) & 0.913 & 3.12(-3) \\
  16.36 & -9.22(-3) & 0.615 & 1.52(-2) & 5.86(-2) & 0.953 & 6.41(-3) \\
  14.32 & -1.03(-2) & 0.586 & 1.86(-2) & 5.87(-2) & 0.895 & 1.00(-2) \\
  12.29 & -1.13(-2) & 0.556 & 2.36(-2) & 5.75(-2) & 0.843 & 1.43(-2) \\
  10.26 & -1.34(-2) & 0.520 & 3.07(-2) & 5.61(-2) & 0.845 & 1.89(-2) \\
  9.243 & -1.47(-2) & 0.502 & 3.59(-2) & 5.56(-2) & 0.791 & 2.07(-2) \\
  8.741 & -1.53(-2) & 0.494 & 3.91(-2) & 5.48(-2) & 0.710 & 2.19(-2) \\
  8.439 & -1.58(-2) & 0.488 & 4.11(-2) & 5.42(-2) & 0.588 & 2.27(-2) \\ \hline
\end{tabular}
\end{center}
\label{table:sequence}
\end{table}

\begin{figure}[t]
\begin{center}
  \includegraphics[width=8cm]{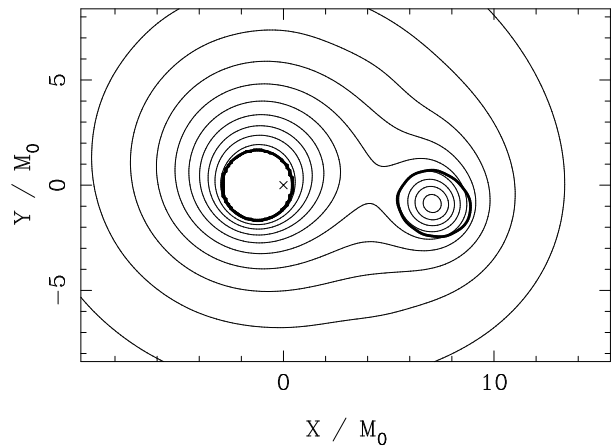}
\end{center}
\caption{Contours of the lapse $\alpha$ in the equatorial plane for
the innermost configuration of the sequence listed in
Table~\ref{table:sequence}.  The thick circle on the left denotes the
excision surface (apparent horizon) of the black hole, while that on the 
right denotes the surface of the neutron star.  The cross ``$\times$'' 
indicates the position of the rotation axis.}
\label{fig:lapse}
\end{figure}

In the following we focus on results for an inspiral sequence of
constant irreducible black hole mass $M_{\rm irr}^{\rm BH}$ and
neutron star baryon rest mass $M_{\rm B}^{\rm NS}$.  In Table
\ref{table:sequence} we list results for a binary of mass ratio
$M_{\rm irr}^{\rm BH}/M_{\rm B}^{\rm NS}=5$ and a neutron star mass
$\bar{M}_{\rm B}^{\rm NS}=0.1$, where the bar denotes non-dimensional
polytropic units $\bar M = \kappa^{-n/2} M$.  At infinite separation,
this neutron star has a compaction $M_{\rm ADM, 0}^{\rm NS}/R_0 =0.0879$,
where $R_0$ is the areal radius of the spherical star in isolation.
In Fig.~\ref{fig:lapse} we also show contours of the lapse $\alpha$
for the innermost configuration of this sequence.  The value of the
lapse everywhere on the black hole apparent horizon is set to be its
Kerr-Schild value there, $2^{-1/2}=0.7071$; its value at the center of
the neutron star is 0.7275.  Note that the center of the neutron star,
as defined by the maximum of the enthalpy, does not coincide with the
minimum point of the lapse inside the star.  These configurations are
the first fully relativistic QE models of BHNS binaries that do not
assume an extreme mass ratio and employ equilibrium boundary
conditions to model the black hole.

In the table we list the fractional binding energy $E_{\rm b}/M_0
\equiv M_{\rm ADM}/M_0 - 1$, total angular momentum $J$, orbital
angular velocity $\Omega$, maximum of the density parameter $q_{\rm
max} = (P/\rho_0)_{\rm max}$, minimum of the mass-shedding indicator
$\chi \equiv (\partial (\ln h)/\partial r)_{\rm eq}/ (\partial (\ln
h)/\partial r)_{\rm pole}$ \cite{TanigBFS05,GourgGTMB01}, and
fractional difference between the Arnowitt-Deser-Misner
(ADM) mass and Komar mass $\delta M
\equiv | 1 - M_{\rm Kom}/M_{\rm ADM} |$. Quantities are tabulated
as functions of the coordinate separation between the center of the
black hole
and the point of maximum baryon rest-mass density in the neutron star.
Here, $M_0=M_{\rm irr}^{\rm BH}+M_{\rm ADM,0}^{\rm NS}$ is the ADM
mass of the binary system at infinite orbital separation, i.e., the
sum of the irreducible mass of the isolated black hole and the ADM
mass of an isolated neutron star with the same baryon rest-mass.  For
an isolated Schwarzschild black hole, the ADM mass is the same as the
irreducible mass.

We compared our values for the angular momentum with those from
third-order post-Newtonian (3PN) approximations \cite{Blanc02} and
found agreement to within about 5\% for close binaries, and better
agreement for larger separations.  In agreement with the 3PN results
we do not see any indication of a turning point in the angular
momentum, meaning that the tabulated sequence does not exhibit an
innermost stable circular orbit, hence the binary orbits are all
stable.

The quantity $\chi_{\rm min}$ is defined by the minimum of the
indicator $\chi$ which compares the gradient of $\ln h$ at the pole
with that on the equator.  For spherical stars at infinite separation
we have $\chi_{\rm min} = 1$, while $\chi_{\rm min} = 0$ indicates the
formation of a cusp and hence tidal breakup.  As we have discussed in
\cite{TanigBFS05}, spectral methods no longer converge in the presence
of discontinuities, so that our sequence terminates before reaching
$\chi_{\rm min} = 0$.  However, extrapolating from the last three data
points we estimate that the star will be tidally disrupted when
$\Omega M_0 \approx 0.046$.  This value agrees with those estimated
via the approximate relativistic expansion of \cite{IshiiSM05}
($\Omega M_0 \approx 0.043$) and the purely Newtonian models of
\cite{UryuE99} ($\Omega M_0 \approx 0.046$).

Equality between the ADM mass and Komar mass is equivalent to
satisfying a relativistic virial theorem and indicates that the system
is stationary (cf. \cite{GourgGB02}).  In our calculation we do not
impose this equality to construct the sequence, but instead evaluate
mass difference as a diagnostic.  For decreasing separation the
fractional difference between the two masses increases to over 2\% for
our innermost configuration.  This clearly indicates that our closest
models are not in perfect equilibrium.  The resulting small, but
finite, systematic mass difference has a large effect on the binding
energy, which is also computed as the small difference between much
larger masses.

We speculate that the differences between the ADM and Komar masses
could be caused by our choice of the background geometry.  Our choice
of a Kerr-Schild background metric is motivated by our requirement
that it correctly reduce
to the exact solution for a spinning black hole in the limit of large
separation (although here we only treat nonspinning holes).
Also, in our coordinates, the lapse remains positive on
the horizon (``horizon penetration''), which is necessary when
computing $\tilde A_{ij}$ from Eq. (\ref{Aij}).  In \cite{TanigBFS05} we
compared with a flat background and found that the choice of the
background has a small but non-vanishing effect on the physical
properties of the resulting binary configurations (see also the
discussion of non-maximally sliced black hole binaries in
\cite{CookP04}), motivating our speculation that our choice here may
result in the small but systematic deviation from perfect equilibrium.
Our configurations are solutions to the constraint equations, and are
hence adequate initial data for dynamical simulations of BHNS binaries
(our main motivation here).  We are currently experimenting with other
background solutions to find better approximations to
quasiequilibrium.  However, this discrepancy in mass also heightens
interest in the recent ``wave-less'' formulation of the initial value
problem that is based on the equality of the ADM and Komar masses and
avoids the need to choose a background geometry altogether
\cite{ShibaUF04,UryuLFGS05}.

Finally, we turn our attention to the validity of QE configurations in
circular orbit as initial data for dynamical simulations. The
assumption of circular orbits and an associated helical Killing vector
for relativistic binaries is an approximation, since the emission of
gravitational waves leads to orbital decay.  This approximation breaks
down at a certain binary separation when the inspiral can no longer be
ignored.  We can quantify the departure from true QE by comparing the
time-scale of the orbital period with that of the orbital decay driven
by the emission of gravitational waves.  To lowest order we can
estimating the ratio between these two timescales with the help of the
quadrupole formula for Newtonian point-masses, which yields
\begin{equation}
{t_{\rm orb} \over t_{\rm GW}} \simeq 0.21 \Bigl(
{d_{\rm min} \over d} \Bigr)^{5/2} 
 \Bigl( {\nu \over 0.135} \Bigr). \label{eq:ratio}
\end{equation}
Here $\nu \equiv M^{\rm BH} M^{\rm NS}/(M^{\rm BH} +M^{\rm NS})^2$ 
and $d_{\rm min}$ denotes the closest binary
separation we computed in this paper (the value of the last line in
Table \ref{table:sequence}).  It is reasonable to approximate the
binary orbit as circular as long as the ratio $t_{\rm orb}/t_{\rm GW}$
is significantly smaller than unity (compare the discussion in
\cite{DuezBSSU02}).  For $d = d_{\rm min}$, we have $t_{\rm orb} \approx
0.2 t_{\rm GW}$, and for larger separations the ratio $t_{\rm
  orb}/t_{\rm GW}$ falls off with $d^{-5/2}$.  For these separations
it is therefore reasonable to neglect the inspiral and construct
binaries in circular orbits and in the presence of a helical Killing
vector.


In summary, we compute sequences of BHNS binaries 
with comparable mass companions.
We solve the constraint equations of general relativity in the
conformal thin-sandwich decomposition, subject to equilibrium black
hole boundary conditions, together with the relativistic equations for
hydrodynamic equilibrium in a stationary spacetime.  We construct
irrotational binaries, adopt a polytropic equation of state for the
neutron star, and choose the background geometry to be a Schwarzschild
black hole expressed in Kerr-Schild coordinates. As an example, we
present results for a binary of mass ratio
$M_{\rm irr}^{\rm BH}/M_{\rm B}^{\rm NS}=5$ and neutron star of
compaction $M_{\rm ADM, 0}^{\rm NS}/R_0=0.0879$.  To the best of our
knowledge, these
are the first models of quasiequilibrium, circular orbit,
relativistic BHNS binaries with companions of comparable mass.


J.A.F. is supported by NSF Grant AST-0401533.  This paper was
supported in part by NSF Grants PHY-0205155 and PHY-0345151, and NASA
Grant NNG04GK54G, to University of Illinois at Urbana-Champaign, and
NSF Grant PHY-0456917 to Bowdoin College.


\end{document}